\documentclass[runningheads]{llncs}

\AtBeginDocument{%
  }
\usepackage[inkscapelatex=false]{svg}
\usepackage{tabularray}
\usepackage{longtable}
\usepackage{booktabs}
\usepackage{tabularx}   
\usepackage{ragged2e}   
\usepackage{subcaption}
\usepackage[T1]{fontenc}
\UseTblrLibrary{diagbox}
\usepackage{float}
\usepackage[most]{tcolorbox}
\usepackage{url}
\usepackage{relsize}
\usepackage{marvosym}

\usepackage[resetlabels,labeled]{multibib}
\newcites{D}{ModelCard and DataSheet References}

\newsavebox{\trashbox}

\begin{document}

\title{From Machine Learning Documentation to Requirements: Bridging Processes with Requirements Languages}
\titlerunning{Machine Learning Documentation to Requirements}

\author{Yi Peng\textsuperscript{(\Letter)}\orcidID{0009-0000-2727-8000} \and Hans-Martin Heyn\orcidID{0000-0002-2427-6875} \and Jennifer Horkoff\orcidID{0000-0002-2019-5277}}
\institute{University of Gothenburg \& Chalmers University of Technology, Sweden \\
 \email{\{yi.peng, hans-martin.heyn, jennifer.horkoff\}@gu.se}}


\authorrunning{Y. Peng et al.}


\maketitle
\begin{abstract}
\vspace{-0.5cm}
In software engineering processes for machine learning (ML)-enabled systems, integrating and verifying ML components is a major challenge. A prerequisite is the specification of ML component requirements, including models and data, an area where traditional requirements engineering (RE) processes face new obstacles.
An underexplored source of RE-relevant information in this context is ML documentation such as ModelCards and DataSheets.
However, it is uncertain to what extent RE-relevant information can be extracted from these documents. 
This study first investigates the amount and nature of RE-relevant information in 20 publicly available ModelCards and DataSheets. 
We show that these documents contain a significant amount of potentially RE-relevant information.
Next, we evaluate how effectively three established RE representations (EARS, Rupp's template, and Volere) can structure this knowledge into requirements.
Our results demonstrate that there is a pathway to transform ML-specific knowledge into structured requirements, incorporating ML documentation in software engineering processes for ML systems. 

\end{abstract}





\vspace{-0.5cm}
\section{Introduction}
\vspace{-0.2cm}
Machine learning (ML)-enabled systems are increasingly used across a wide variety of domains, including healthcare, automotive, and manufacturing~\cite{Iqbal2022AssuranceOM}. 
However, their non-deterministic behavior and heavy reliance on data complicates software engineering (SE) processes, especially in managing customer expectations, data quality, and ensuring transparency in system design~\cite{alves2023status}.
Requirements Engineering (RE) plays a central role in addressing these challenges by translating stakeholder needs through clear and actionable requirements~\cite{Habibullah2021NonfunctionalRF}. With ML components as part of complex software systems, there is a need to re-consider where requirements-related information comes from, and what types of new artifacts or boundary objects we should be considering as part of SE processes for ML systems~\cite{kasauli2020charting}.  
Defining and integrating processes to capture model and data requirements, which cover aspects such as robustness, interpretability, fairness, and data quality~\cite{pradhan2024identifying}, remains a challenge in SE for ML systems~\cite{letier2025obstacle}.\par

In parallel, the ML community has introduced documentation artifacts such as ModelCards~\cite{mitchell2019model} and DataSheets~\cite{Gebru2021Datasheets} to document component-level information. 
These artifacts are particularly popular in open-source platforms such as Hugging Face
(\url{https://huggingface.co/models}), yet they remain loosely structured and inconsistently adopted which limits their integration into software development processes~\cite{bhat2023aspirations}. Nonetheless, we argue that these artifacts could be a potential source of information for RE because they document information about intended use, performance limitations, and biases, which directly relate to requirements, including functional requirements (FRs), non-functional requirements (NFRs), and constraints. Structuring this information through RE languages may be a key step towards integrating these informal artifacts into SE for ML systems processes.\par

The overall goal of this study is to assess the potential of ModelCards and DataSheets to serve as a bridge between informal ML component descriptions and more structured SE processes. 
To determine if these artifacts can serve this role, we must first validate that they contain valuable requirements-relevant information in practice. This step is addressed by research question (RQ) 1:\par

\begin{description}
\item[RQ1] How much requirements-relevant information is contained in ModelCards and DataSheets?
    \begin{description}
        \item[RQ1.1] What are the issues in how this information is documented from an RE perspective? 
    \end{description}
\end{description}
If the presence of this information is confirmed, the next step is to evaluate whether existing RE languages are suitable for structuring it. RQ2 is formulated to investigate this:\par
\begin{description} 
    \item[RQ2] How effective  are existing requirements representation languages (e.g., EARS, Rupp’s Template, Volere) in capturing requirements-relevant information from ModelCards and DataSheets?
    \begin{description}
        \item[RQ2.1] What types of requirements-relevant information are not well captured by these techniques in the context of ModelCards and DataSheets?
    \end{description} 
\end{description}

\noindent We investigated these questions in two steps: First, we analyze a sample of 20 diverse ModelCards and DataSheet, revealing that these artifacts contain significant RE-relevant information. 
However, they also exhibit redundancies and inconsistent levels of detail, which may stem from the varied adoption of these documentation practices.
Second, we evaluate how well three well-known RE languages \textemdash EARS~\cite{mavin2009ears}, Rupp’s Template~\cite{rupp2007requirements}, and Volere~\cite{robertson1999volere} \textemdash capture this RE-relevant content, finding that while EARS and Rupp's Template can extract much of the RE-relevant information, only Volere captures implementation-level details relevant for integrating ML models and data into broader system specifications. Overall, our findings contribute to the integration of ML documentation artifacts into the software engineering life cycle, aiming to improve communication between ML and SE teams and enhance the engineering of successful ML-systems.\par
The paper is organized as follows: Sec.~\ref{sec:background} presents background and related work, with methodology in Sec.~\ref{sec:methodology}. We report the results in Sec.~\ref{sec:results} and discuss implications, future work, and threats in Sec.~\ref{sec:discussion}. Sec.~\ref{sec:conclusion} concludes this paper.

\section{Background and Related Work}
\label{sec:background}

\subsection{RE Languages and Templates} \label{subsec:RElangs}
\begin{figure}[hbt]
\vspace{-0.1cm}
    \centering
    \begin{subfigure}[b]{0.49\linewidth} 
        \centering 
        \includesvg[width=\linewidth]{EARS_template.svg}
        \caption*{(a)} 
        \label{fig:ears} 
    \end{subfigure}
    \hfill 
    \begin{subfigure}[b]{0.49\linewidth} 
        \centering 
        \resizebox{\linewidth}{4cm}{
        \includegraphics[width=\linewidth]{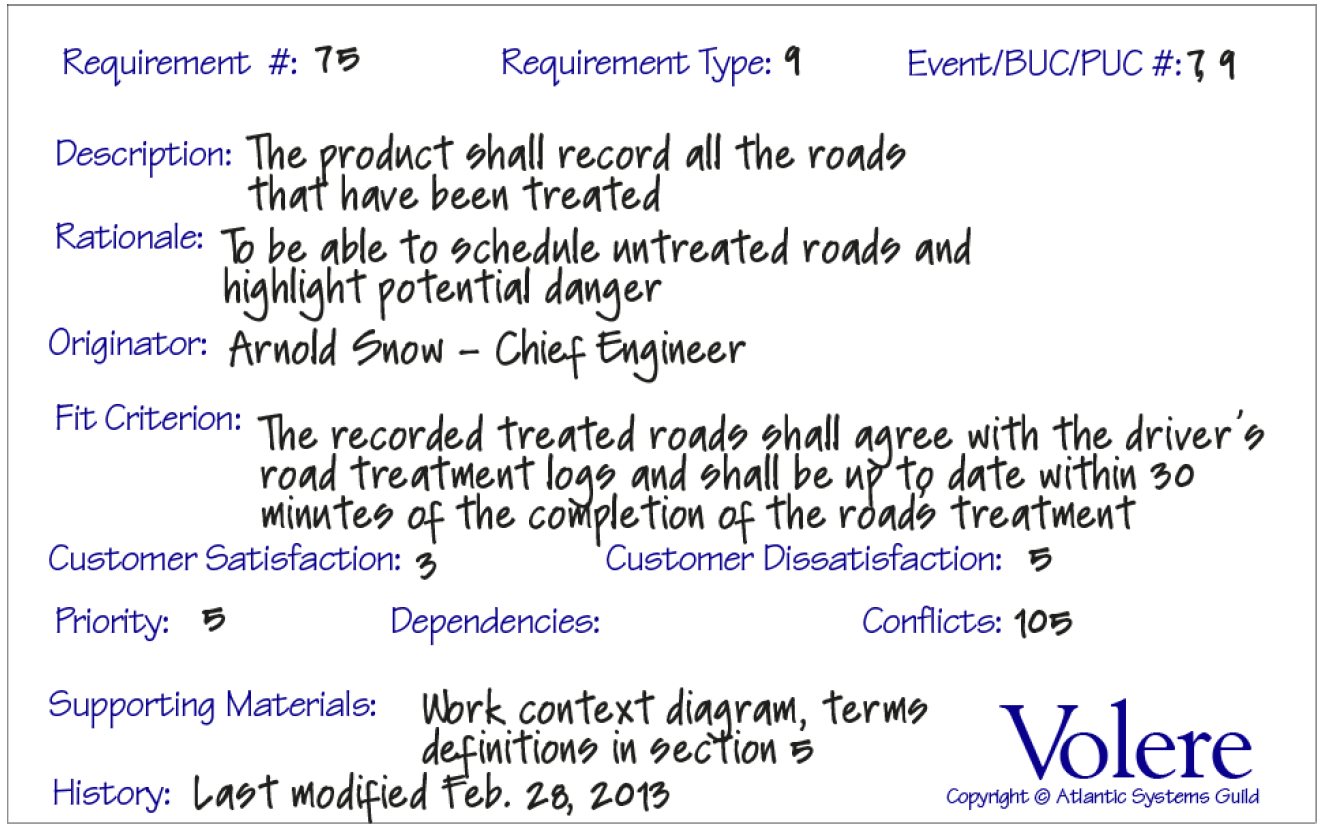}
    }
        \caption*{(c)} 
        \label{fig:rupp} 
    \end{subfigure}

    \vspace{\baselineskip} 

    \vspace{-0.5cm}
    \begin{subfigure}[b]{0.70\linewidth} 
        \centering 
        \includesvg[width=\linewidth]{Rupps_template.svg}
        \caption*{(b)} 
        \label{fig:volere} 
    \end{subfigure}

    \caption{(a) EARS~\cite{mavin_ears}, (b) Rupp's template~\cite{mazo2020towards} and (c) Volere~\cite{robertson2012mastering}.}
    \label{fig:RE languages}
    \vspace{-0.5cm} 
\end{figure}

Structured RE languages and templates enhance clarity, traceability, and consistency in requirements documentation. 
Among them, EARS, Rupp's template, and Volere are widely adopted~\cite{grosser2024benchmarking,gimenez2021inter}. 
Fig.~\ref{fig:RE languages} shows the core structures of EARS, Rupp's Template and Volere.

\textbf{EARS} is a lightweight, structured natural language format, applied successfully in industry~\cite{mavin2009ears}. 
It uses patterns such as ubiquitous, event-driven, state-driven, optional feature, and unwanted behavior to reduce ambiguity and improve readability~\cite{Gregory2011EasyE}.\par
\textbf{Rupp's Template}~\cite{rupp2007requirements}, established by the International Requirements Engineering Institute (IREB) as a de facto standard for the syntactic specification of system requirements~\cite{mazo2020towards}, specifies nine patterns: three for FRs, three for NFRs, and three for conditional clauses. 
FRs are grouped under FunctionMASTER, classified by activity type: system, user interaction, and interface. 
NFRs are expressed using PropertyMASTER, EnvironmentMASTER, and ProcessMASTER. 
Conditional logic is supported by LogicMASTER (“if”), EventMASTER (“as soon as”), and TimeMASTER (“as long as”). 
The template encourages fine-grained specification using modal verbs and optional details~\cite{grosser2024benchmarking}.\par
\textbf{Volere}~\cite{Robertson2019Volere} focuses on traceability of requirements and structured metadata (e.g., rationale, acceptance criteria). 
It categorizes NFRs (33 types) and constraints (8 types), and includes sections on external factors (e.g., assumptions, risks). 
With over 20,000 downloads of its online template materials reported on its official website, the Volere template demonstrates significant practical use.\par 
These three RE languages were selected for their popularity and distinct strengths: 
EARS provides a lightweight, structured syntax for atomic requirements; 
Rupp’s template offers a systematic approach for capturing variability and fine-grained system behavior, and Volere ensures traceability and completeness. 
Together, they offer a comprehensive lens to assess how information in ML documentation can be utilized as a source of requirements.

\subsection{RE for ML Systems}
Vogelsang and Borg \cite{vogelsang2019requirements} were among the first to define characteristics and challenges unique to RE for ML-based systems by interviewing data scientists, observing that the development paradigm for ML requires rethinking of RE.
Since then, frameworks such as RE for human-centered AI~\cite{ahmad2023requirementsFramework}, perspective-driven RE~\cite{villamizar2024identifying}, as well as extensions of goal-orientated RE~\cite{barrera2024extension} have emerged.
While these approaches guide requirement elicitation and modeling, they tend to remain high-level and lack concrete guidance for specifying ML requirements.
A recent review by Habiba et al.~\cite{habiba2024mature} highlights remaining difficulties in concretely specifying low-level requirements, insufficient documentation guidelines, and the need to adapt existing RE practices for ML-based systems.\par
Several proposed RE for ML frameworks have utilized RE representation such as EARS, and UML to capture FRs for ML systems~\cite{yang2024rm4ml,al2022resam}. Dedicated templates~\cite{Bajraktari2024DocumentationON} and Volere~\cite{habibullah2024framework} have been applied to specify NFRs. While these approaches provide valuable frameworks for documenting ML system requirements, they focus on specific types of requirements such as FRs, NFRs or quality requirements, and limited dimensions of data requirements, which may not fully capture the breadth of information needed for ML system development.
We argue that ML-specific documentation artifacts like ModelCards and DataSheets, which are created during the actual development of ML models and datasets, are a potential source of such information. They offer broad information spanning use cases (FRs), fairness considerations (NFRs), environmental assumptions, and hardware constraints. While related work has mined less structured sources like GitHub README files for requirements~\cite{portugal2017gh4re}, we focus on ModelCards and DataSheets for ML-specific requirements, against which we evaluate the applicability and coverage of existing RE languages.

\subsection{ModelCards and DataSheets}
\begin{figure}[bht]
    \centering
    \begin{subfigure}[b]{0.45\linewidth} 
        \centering
        \includegraphics[width=\linewidth]{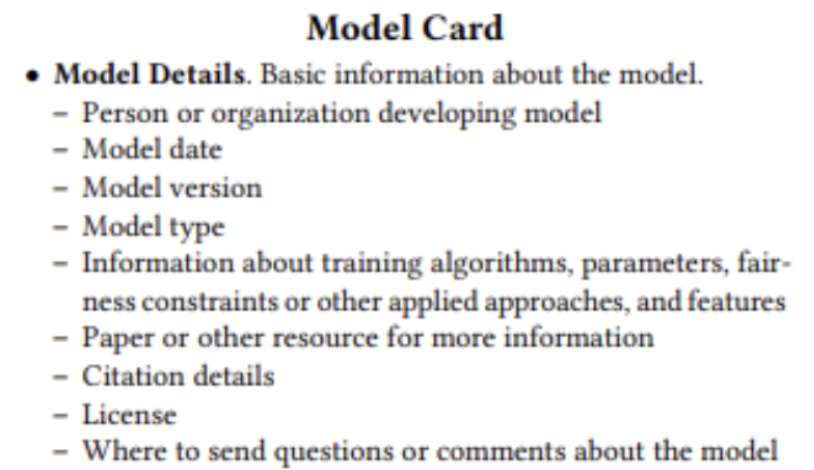}
        \caption{ModelCards template excerpt.}
        \label{fig:modelcardExcerpt}
    \end{subfigure}
    \hfill
    \begin{subfigure}[b]{0.45\linewidth} 
        \centering
        \includegraphics[width=\linewidth]{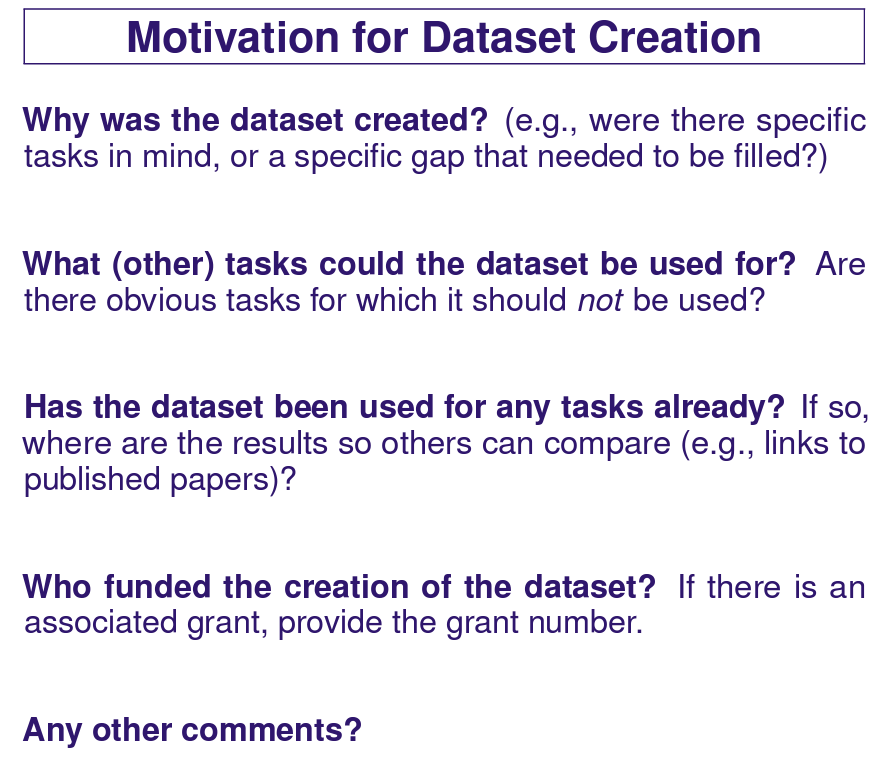}
        \caption{Datasheet template excerpt.}
        \label{fig:datasheetExcerpt}
    \end{subfigure}
    \vspace{-0.2cm}
    \caption{Examples of template excerpts from ModelCards~\cite{mitchell2019model} and DataSheet~\cite{Gebru2021Datasheets}.}
    \label{fig:templateExcerpts}
    \vspace{-0.6cm}
\end{figure}

ML documentation artifacts, including ModelCards~\cite{mitchell2019model} and DataSheets~\cite{Gebru2021Datasheets}, exemplified in Fig.~\ref{fig:templateExcerpts}, aim to support transparency, accountability, and ethical considerations in ML development. 
ModelCards document a model's performance, intended use, and limitations. 
DataSheets aim to ease assessment of a dataset's suitability by describing properties such as motivation, composition, collection processes, and potential biases.
Their popularity is evidenced by high citation counts (2,000+ each) and ModelCard adoption in the open-source platform Hugging Face. 

Despite their popularity, recent studies show that ML documentation artifacts are often inconsistent~\cite{bhat2023aspirations}, incomplete~\cite{heger2022understanding}, and lack clear purpose or organizational incentives~\cite{chang2022understanding}. As such, it can be difficult to integrate these artifacts into established SE processes.
However, they describe functionality, quality, assumptions, and constraints \textemdash all potentially requirements-relevant information. To the best of our knowledge, this is the first study analyzing and utilizing these ML documentation from a RE perspective in SE for ML systems processes.

\section{Methodology}
\label{sec:methodology}
\vspace{-0.3cm}
This section describes the selection of data, the evaluation of requirements-relevant content, the extraction of structured requirements, and the validation procedures. 
All raw and processed data are available in our online repository\footnote{\url{https://doi.org/10.6084/m9.figshare.28564058.v1}}.
\vspace{-0.2cm}

\subsection{Data Selection}
\vspace{-0.2cm}
We selected ten ModelCards and ten DataSheets, as listed in Table~\ref{tab:modelCardList}, with sources referenced using a ``D'' and citations included in our online material. Because our purpose was to evaluate the potential presence of RE-relevant information in ModelCards and DataSheets, we aimed to find high-quality examples of each.
We found these documents using Google Scholar by searching for studies that cited the original papers on ModelCards and DataSheets~\cite{mitchell2019model,Gebru2021Datasheets}. 
This method assumes that authors citing these papers are more likely to create ML documentation that follow the original intended purpose and structure.\par
Our selection aimed to ensure: 1) \emph{diversity}, by including examples which can be used in different fields like healthcare, security, and commerce; 2) \emph{variety}, by sourcing documents from both academic projects and industry work; and 3) \emph{paired documents}, including both ModelCard and a DataSheet for the same ML system when available (e.g., MC2,DS2 and MC3,DS3) in order to see how they document an ML artifact together. 
The final dataset includes over 1,200 analyzed sentences. 
While publicly accessible repositories of ModelCards and DataSheets exist (e.g., downloadable via API from the Hugging Face website, and in curated research datasets~\cite{zilka2022survey,pepe2024hugging}), we did not sample directly from these repositories because they are often compiled for specific analytical goals (e.g., criminal justice~\cite{zilka2022survey} and transformer models only~\cite{pepe2024hugging}) that does not align with our desire for diversity across various contexts.

{\scriptsize
\vspace{-0.3cm}
\begin{longtblr}[
    caption = {List of ModelCards and DataSheets (Citations in supplementary material)},
    label = {tab:modelCardList}
]{
    width = \linewidth,
    colspec = {Q[90]Q[150]Q[485]Q[70]},
    vlines,
    hline{1-2,12-13,23} = {-}{},
}
Identifier & ModelCard & Description & Source\\
MC1 & FaceDetect-IR & The model detects one or more faces in the given image/ video. & \citeD{nvidia_facedetectir} \\
MC2 & SAM & Model for any prompt-based image segmentation task. & \citeD{kirillov2023segment} \\
MC3 & DynaSent & Sentiment analysis of the texts of product and service reviews. & \citeD{potts2020dynasent} \\
MC4 & LLAMA 2 & Collection of pretrained and fine-tuned large language models. & \citeD{touvron2023llama} \\
MC5 & Summariza-tion model & Model for summarizing text. & \citeD{stiennon2020learning} \\
MC6 & InstructGPT & A GPT-style language model fine tuned to follow instructions from human feedback. & \citeD{ouyang2022training} \\
MC7 & RoentGen & A model pre-trained on pairs of natural images and text descriptors to generate synthetic chest X-ray images. & \citeD{bluethgen2024vision} \\
MC8 & Seamless-M4T & Multilingual and multimodal translation models. & \citeD{barrault2023seamlessm4t} \\
MC9 & StoryDALL-E & A model trained for the task of Story Visualization. & \citeD{maharana2022storydall} \\
MC10 & BRAVE-NET & Model for arterial brain vessel segmentation. & \citeD{hilbert2020brave} \\
Identifier & DataSheet & Description & Source\\
DS1 & CheXpert & Public dataset for chest radiograph interpretation. & \citeD{irvin2019chexpert}\\
DS2 & SA-1B & Dataset of images for image segmentation. & \citeD{kirillov2023segment}\\
DS3 & DynaSent & English-language benchmark task for ternary (positive/negative/neutral) sentiment analysis & \citeD{potts2020dynasent}\\
DS4 & Movie review polarity & Movie reviews extracted from newsgroup postings together with a sentiment polarity rating. & \citeD{Gebru2021Datasheets}\\
DS5 & Meta-album & Multi-domain meta-dataset for few-shot image classification & \citeD{ullah2022meta}\\
DS6 & SVIB & Test-bed to evaluate the systematic generalization ability of visual imagination models. & \citeD{kim2024imagine}\\
DS7 & Youtube ASL & Training data for ASL to English machine translation. & \citeD{uthus2024youtube}\\
DS8 & Change Event Dataset & Dataset for developing systems that can automatically detect change events in satellite imagery. & \citeD{mall2022change}\\
DS9 & SituatedQA & Information seeking questions that is annotated for its temporal or geographical dependence. & \citeD{zhang2021situatedqa}\\
DS10 & Egoschema & Diagnostic benchmark for assessing long-form video-language understanding capabilities. & \citeD{mangalam2023egoschema} 
\end{longtblr}
\vspace{-0.6cm}
}

\subsection{Identification of Requirements-relevant Information}  
We followed a \emph{deductive thematic analysis} approach described by Braun and Clarke~\cite{braun2021thematic}. 
Our coding scheme is based on ISO/IEC/IEEE 29148:2018~\cite{ISO29148-2018}.
The aim of the chosen coding scheme is to identify information within ModelCards and DataSheets that aligns with what the standard considers to be requirement-relevant or essential supporting information for requirements. We therefore derived deductive codes from these sections of the ISO standard: 
\vspace{-0.2cm}
\begin{description} 
    \item \textbf{Clause 3.1:} Terms and definitions: the definition of a ``requirement'' itself.
    \item \textbf{Clause 5.2:} Requirements fundamentals: how requirements should be formed, what qualities they should possess, and what associated information gives them context.
    \item \textbf{Clause 9:} Information item content: details general content for business, system and software requirements specification.
    \item \textbf{Annex A:} defines system operational concept, providing guidance on user-oriented descriptions of system characteristics and operational scenarios.
\end{description}
We excluded clauses focusing on the RE process itself (e.g., Clause 6) or on conformance to the standard (e.g., Clause 4). 
Table~\ref{tab:REinfo} presents the deductive coding criteria, with a more detailed version showing examples from the chosen ML documents, available in the supplementary material.

{\scriptsize
\vspace{-0.4cm}
\begin{longtblr}[
  caption = {Requirements-relevant Info Deductive Codes Criterion},
  label = {tab:REinfo}
]{
  width = \linewidth,
  colspec = {Q[550]Q[450]},
  vlines,
  hline{1-2,10} = {-}{},
} 
Requirements-relevant Info. & Requirements-irrelevant Info.\\ 
Explicitly states or clearly implies a need, purpose, goal, or intended use of the model/dataset. & Is purely descriptive of the ML model's architecture or dataset creation process \textit{without} direct implications for its use, performance, capabilities, or constraints.\\
Describes a specific capability, function, or task the model/dataset performs or supports. & Provides general ML or domain knowledge not specific to \textit{this} model/dataset's behavior or characteristics.\\
Specifies performance characteristics or quality attributes, especially if quantitative or verifiable. & Consists of aspirational statements, highly speculative ``future work,'' or desired features not currently implemented or guaranteed.\\
Defines operational conditions, constraints, or limitations and negative impacts (e.g., ethical concerns like bias, safety considerations, security vulnerabilities) for the model/dataset's use.  & Is purely bibliographic, author names, funding acknowledgments, unless imposing a usage constraint.\\
 Describes assumptions or dependencies crucial for the model/dataset's correct or intended functioning. & Is raw data within a dataset (e.g., pixel values) as opposed to metadata \textit{about} the dataset or statements regarding its use, quality, or format.\\
 Specifies interface details for interaction with the model/dataset. & \\
Provides information related to user characteristics or the intended operational environment/context of use. & \\
 Details evaluation methods, or metrics used, implying how the model/dataset should be assessed or what constitutes acceptable performance.  & 
\end{longtblr}
\vspace{-0.8cm}
}

\subsection{Requirements Extraction Process} 
Once requirements-relevant information was identified, we manually extracted structured requirements using the three selected representation languages EARS, Rupp's Template, and Volere. 
Although we followed central tenants of thematic analysis from Braun and Clarke~\cite{braun2021thematic}, this process was more-so extraction than qualitative coding,
similar to previous work in RE~\cite{grosser2024benchmarking}.

The first author read the text from the ML documentation artifact, then selected the best-fitting structure from a given template for each requirement-relevant sentence (e.g., one of the five EARS types), and finally mapped the information into structural components.
This mapping step involved identifying the subject, formulating the primary assertion (e.g., \texttt{<system response>} in EARS, or `Description' in Volere), and populating the template's specific slots or attributes (e.g., EARS preconditions/triggers, Rupp's template conditional clauses) with corresponding contextual details. 
Information for some of Volere's meta fields (e.g., `Originator', `Customer Satisfaction') was often missing and therefore excluded.
External factors (e.g., `Goals of Project', `Risks') in Volere were linked using the `Dependencies' field. Examples of extracted requirements appear in Sec.~\ref{sec:results}.
During this process, we also documented recurring patterns of documentation issues from an RE perspective.

\subsection{Validation} We iteratively validated the data for each requirements representation language. After extraction by the first author, 
the other authors independently reviewed a 10\% random sample, following the guidance by Lombard et al.~\cite{lombard2002content}.  
A total of two iterations were conducted. The initial iteration stopped after extraction issues were found with EARS and Rupp's template, finding inconsistency in judging what information was requirements-relevant (author \#2: 19 disagreements out of 325 requirements extracted from the 10\% sample, author \#3: 35/317). 
To resolve this, we developed our deductive coding criteria (Table~\ref{tab:REinfo}) to create a shared standard. In the second iteration, we reviewed a different 10\% sample of all three representations and disagreements shifted to how to best represent certain types of information, particularly contextual details and limitations, within the formats of our chosen RE languages (author \#2: 10/331 requirements extracted from the new 10\% sample, author \#3: 17/331). 
All disagreements were resolved through team discussions with the help of the deductive coding criteria and guidelines for the specific RE language. 
The first author then adjusted the entire dataset based on the results of the discussions. 

\subsection{Quantitative Data Analysis}
To answer RQ1, we counted all sentences identified as requirements-relevant. Partially relevant sentences were counted as 0.5. For example in DS3: ``[Is there an erratum] Not at present, but we will create one as necessary and update this section at that time.'' The first half of the sentence describes an operational limitation that is relevant according to our criterion, but the latter half is a desired feature not guaranteed, and considered irrelevant from our criterion.
For RQ2, we evaluated how well each RE language captured the identified requirement-relevant information. 
The requirements representations were assessed by computing the percentage of relevant sentences they captured, using the same 0.5 point rule for partially relevant sentences. 
Volere was also evaluated by computing the percentages of relevant sentences it captured as external factor statements.

\section{Evaluation results}
\label{sec:results}

\subsection{\textbf{RQ1}: Requirements Relevant Information in ModelCards and DataSheets}
The total number of sentences per documentation artifact, the number of requirements relevant sentences, and the corresponding percentage of requirements relevant information is shown in Table~\ref{tab:reqrelevant}.
{
\scriptsize
\begin{longtblr}[
  caption = {Statistics for Information Considered Requirements-relevant},
  label = {tab:reqrelevant}
]{
  width = \linewidth,
  colspec = {Q[304]Q[67]Q[60]Q[67]Q[60]Q[60]Q[60]Q[60]Q[60]Q[60]Q[71]},
  vline{1-2,12} = {-}{},
  hline{1-2,5-6,9} = {-}{},
}
                                  & MC1   & MC2  & MC3   & MC4  & MC5  & MC6  & MC7  & MC8  & MC9  & MC10 \\
\# of sentences (S)               & 69    & 41   & 23    & 30   & 61   & 50   & 25   & 42   & 27   & 32   \\
\# of relevant sentences (RS)     & 53    & 37   & 16    & 25   & 58   & 45   & 18   & 37   & 24.5 & 24   \\
\% of relevant information (RS/S) & 0.77  & 0.90 & 0.70  & 0.83 & 0.95 & 0.90 & 0.72 & 0.88 & 0.91 & 0.75 \\
                                  & DS1   & DS2  & DS3   & DS4  & DS5  & DS6  & DS7  & DS8  & DS9  & DS10 \\
\# of sentences (S)               & 206   & 89   & 129   & 74   & 75   & 70   & 67   & 108  & 69   & 99   \\
\# of relevant sentences (RS)     & 178.5 & 79   & 109.5 & 63   & 69   & 61   & 56.5 & 99.5 & 63   & 96   \\
\% of relevant information (RS/S) & 0.87  & 0.89 & 0.85  & 0.85 & 0.92 & 0.87 & 0.84 & 0.92 & 0.91 & 0.87 
\end{longtblr}
\vspace{-0.2cm}
}
\noindent On average, 83\% of ModelCard sentences and 88\% of DataSheet sentences were found to be potentially relevant.
ModelCards ranged from 70\% to 95\%, while DataSheets were more consistent, between 84\% and 92\%.\par

\paragraph{\textbf{RQ1.1:} Issues with ML documentation.} 
We observed several issues with how information is documented as part of ML Documentation.
First, redundancy: the same information was often expressed multiple times in different phrasings or sections.
For example, DS1 repeatedly describes the need for radiologist-labeled data using overlapping statements: 
{\footnotesize \textit{``The dataset shall have validation and test sets labeled by multiple board-certified radiologists providing a strong ground truth to evaluate models''}, \textit{``The dataset labels shall have been evaluated against labels manually extracted by board-certified radiologists''}, and \textit{``The dataset shall have strong radiologist-annotated ground truth''}}. 
Second, variation in granularity: Some sections offer high-level, abstract descriptions, while others contain low-level implementation details that are more design decisions than requirements.\par
\textbf{Answer to RQ1.1:} From an RE perspective, the primary issues with how information is documented in the chosen ModelCards and DataSheets are the redundancy of information and inconsistent level of granularity.\par
\textbf{Answer to RQ1:} Within our chosen samples, both ModelCards and Data-Sheets are dense with requirements-relevant information (averaging over 80\% of sentences), making them a rich source for RE. However, the quality of these documentation may be hampered by redundancy and inconsistent granularity.\par

\subsection{\textbf{RQ2:} Capturing Requirements Relevant Information using Three RE Languages}
Table~\ref{tab:stats} summarizes to what degree the chosen RE languages EARS, Rupp's template, and Volere capture the identified requirements-relevant information from ML documentation artifacts.

{\scriptsize
\vspace{-0.3cm}
\begin{longtblr}[
  caption = {Statistics for Information Captured Using RE Templates(E = EARS, R = Rupp's tmp., V = Volere)},
  label = {tab:stats}
]{
  width = \linewidth,
  colspec = {Q[267]Q[63]Q[63]Q[63]Q[63]Q[63]Q[63]Q[63]Q[63]Q[63]Q[79]},
  vline{1-2,12} = {-}{},
  hline{1-2,6-7,11} = {-}{},
}
                                            & MC1  & MC2  & MC3  & MC4  & MC5  & MC6  & MC7  & MC8  & MC9  & MC10 \\
{\% information\\captured in E}       & {0.79} & {0.64} & {0.66} & {0.76} & {0.44} & {0.37} & {0.58} & {0.64} & {0.51} & {0.58} \\
{\% information\\captured in R} & 0.79 & 0.72 & 0.69 & 0.76 & 0.53 & 0.41 & 0.58 & 0.66 & 0.53 & 0.63 \\
{\% info. captured as V reqs.}               & 0.5   & 0.73   & 0.66   & 0.64   & 0.23   & 0.36   & 0.5   & 0.41    & 0.43   & 0.38    \\
{\% info. captured as V ext. factors}           & 0.5   & 0.27    & 0.34    & 0.36    & 0.77   & 0.64   & 0.5    & 0.59    & 0.57   & 0.62   \\
                                            & DS1  & DS2  & DS3  & DS4  & DS5  & DS6  & DS7  & DS8  & DS9  & DS10 \\
{\% information\\captured in E}       & 0.49 & 0.71 & 0.55 & 0.63 & 0.75 & 0.76 & 0.58 & 0.71 & 0.48 & 0.64 \\
{\% of information\\captured in R} & 0.53 & 0.76 & 0.63 & 0.71 & 0.78 & 0.80 & 0.54 & 0.70 & 0.60 & 0.65 \\
{\% info. captured as V reqs.}           & 0.59   & 0.64   & 0.49   & 0.62   & 0.74   & 0.65   & 0.43   & 0.61   & 0.52   & 0.67   \\
{\% info. captured as V ext. factors}        & 0.41   & 0.36   & 0.51   & 0.38   & 0.26   & 0.35   & 0.57   & 0.39   & 0.48   & 0.33 
\end{longtblr}
\vspace{-0.3cm}
}
EARS captured 37\% (MC6) to 79\% (MC1) of the sentences of relevant ML documentation, while Rupp's template captured 41\% (MC6) to 80\% (DS6). 
For ModelCards, both EARS and Rupp's template worked well in capturing input-output specifications, training algorithms, integration interfaces, usage license constraints, primary intended users and use cases, as well as for defining metrics for model evaluation. 
For DataSheets, both templates captured motivation, composition, collection, distribution, and maintenance. 
Examples of requirements captured by EARS and Rupp's template are shown in Table~\ref{tab:combined_examples}. 
Rupp's template outperformed EARS in capturing requirements from ML documentation in seven out of ten ModelCards and in eight out of ten DataSheets, with a difference of up to 12\%(DS9).\par
{
\begin{table}[!htb]
\caption{Examples of requirements captured in EARS and Rupp's template.}
\label{tab:combined_examples}

\begin{minipage}[t]{0.49\textwidth}
    \scriptsize
    \centering
    \caption*{EARS Examples} 
    \label{tab:EARS_requirements_examples_condensed}
    \begin{tabularx}{\linewidth}{@{} p{1.6cm} >{\RaggedRight}X @{}}
    \toprule
    \textbf{Pattern} & \textbf{Captured requirement and original text} \\ 
    \midrule
    Ubiquitous & 
    \textit{The model shall be a 1.3B parameter Transformer model trained with human feedback.} \\
    \multicolumn{2}{p{\linewidth}}{\footnotesize MC5(primary use): This model card details the 1.3 billion parameter Transformer model trained with human feedback.} \\ 
    \midrule
    
    Event-driven & 
    \textit{When the pruned model is selected, the model shall accept 384x240x3 dimension input tensors...} \\
    \multicolumn{2}{p{\linewidth}}{\footnotesize MC1 (input/output specification): The pruned model... accepts 384x240x3 dimension input tensors and output...} \\ 
    \midrule

    State-driven & 
    \textit{While training, the model shall be trained and fine-tuned against a cross-entropy loss function.} \\
    \multicolumn{2}{p{\linewidth}}{\footnotesize MC3(training algorithm): The model parameters are trained/fine-tuned against a cross-entropy loss.} \\ 
    \midrule

    Unwanted behavior & 
    \textit{If Yelp removes data..., the dataset shall remain intact and self-contained.} \\
    \multicolumn{2}{p{\linewidth}}{\footnotesize DS3(composition): ...cannot guarantee that the Yelp Academic Dataset will always be available... Our dataset would remain intact even if this did happen.} \\
    \bottomrule
    \end{tabularx}
\end{minipage}%
\hfill 
\begin{minipage}[t]{0.49\textwidth}
    \scriptsize
    \centering
    \caption*{Rupp's Template Examples} 
    \label{tab:Rupp_examples_condensed}
    \begin{tabularx}{\linewidth}{@{} p{1.6cm} >{\RaggedRight}X @{}}
    \toprule
    \textbf{Pattern} & \textbf{Captured requirement and original text} \\ 
    \midrule
    Function-MASTER (in.~sys.~act.) & 
    \textit{The model shall accelerate and facilitate inspection of brain vessels in clinical practice...} \\
    \multicolumn{2}{p{\linewidth}}{\footnotesize MC10(motivation): ...accelerating and facilitating inspection of brain vessels in clinical practice for diagnosis of vessel...} \\ 
    \midrule

    Function-MASTER (user intera.) & 
    \textit{The model shall provide users who may not have AI development experience with the ability to build and explore language modeling systems...} \\
    \multicolumn{2}{p{\linewidth}}{\footnotesize MC10(integration interfaces): Through the OpenAI API, the model can be used by those who may not have AI development experience...} \\ 
    \midrule

    Logic-MASTER, propertyMASTER & 
    \textit{If an image is with protected health information of any type, the image shall be excluded from the dataset.} \\
    \multicolumn{2}{p{\linewidth}}{\footnotesize DS1(collection): Images that had protected health information (PHI) of any type were excluded from the dataset.} \\ 
    \midrule

    Property-MASTER & 
    \textit{The license for model use shall be CC-BY-NC 4.0 b} \\
    \multicolumn{2}{p{\linewidth}}{\footnotesize MC8(license constraint): Model details: – License: CC-BY-NC 4.0 b} \\
    \bottomrule
    \end{tabularx}
\end{minipage}
\vspace{-0.3cm}
\end{table}
}
Unlike EARS and Rupp's template, Volere also captures supplementary information from ML documentation that, while not strictly requirements, can be considered as external factors relevant to RE. As such, we judge that Volere is able to capture all the requirements relevant information contained in the chosen sample of ML documentation artifacts. This can be seen in Table~\ref{tab:stats} where for each document, the sum of percentages of Volere requirements and external factors is 100\%. The percentage of external factors is quite substantial, and Volere's capability of documenting external factors is particularly useful for capturing information affecting the design implementation and constraints that may influence the ML specification such as license dependencies. 
For example, MC1 includes a legal requirement (license compliance) linked to a supporting fact (user agreement), which can be captured using Volere’s ``Dependencies'' field. \par
\begin{figure} 
  \centering
  \includesvg[width=0.60\linewidth]{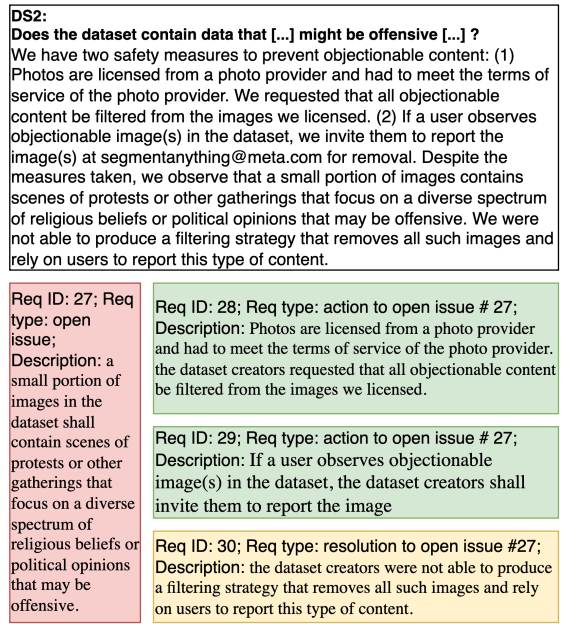}
  \caption{Example of Volere capturing open issues.}
  \label{fig:open issue example}
  \vspace{-0.6cm}
\end{figure}

Volere also allows capturing ``open issues'' of the models and datasets and to associate them with ``actions'' or ``resolutions'' as illustrated in Figure~\ref{fig:open issue example}.
Similarly, Volere is able to document constraints over out-of-scope usage, variables that influence model performance, and limitations from ModelCards. 
From DataSheets, Volere can document limitations in the collection and preprocessing steps, unintended uses of the dataset, and potential risks in the dataset composition.\par

\vspace{-0.3cm}

\paragraph{\textbf{RQ2.1:} Limitations of the RE templates.}
Rupp’s template captured more nuanced content than EARS due to its broader modal vocabulary.
For example, DS2 states that the dataset is ``believed to be more representative''.
This uncertainty is lost in EARS as it only allows ``shall'' as a modal verb, while Rupp's template  allows  ``should'', or ``may'', which both capture this uncertainty.\par

A broader challenge for both EARS and Rupp's template is that
they cannot intuitively express known limitations and constraints which are requirements relevant, such as, for example, MC4’s testing being limited to English, or DS8's change detection algorithm being not 100\% accurate. Although this information could be very relevant to a requirement such as ``the model shall be tested in multiple languages'' and ``the change detection algorithm should be 100\% accurate'', such requirements do not accurately reflect the content of the ModelCard or DataSheet, and therefore is not easily captured by EARS and Rupp's template.\par

Volere's main limitation is its lack of ML-specific requirement types, forcing the conflation of distinct concepts into its generic categories. For example, a constraint on training data sources and a constraint on evaluation metrics were both categorized simply as a ``solution constraint'' (MC3). Similarly, a data quality requirement concerning author diversity (DS4) was also classified as a ``solution constraint'' due to the absence of a dedicated data quality category.\par

\textbf{Answer to RQ2.1:} EARS fails to capture uncertainty due to its rigid vocabulary. Both EARS and Rupp's template struggle to document contextual information such as limitations and constraints. Volere's generic categories conflate ML-specific concepts.

\textbf{Answer to RQ2:} EARS and Rupp's template can capture a majority of the requirements-relevant information while Volere proves more comprehensive: capturing all relevant information by structuring requirements alongside supplementary ``external factors'' such as dependencies, limitations and open issues. EARS is limited in its ability to capture uncertainty, while both EARS and Rupp's template struggle to document contextual information. Volere's effectiveness is limited by a generic typology that conflates ML-specific concepts.\par

\section{Discussion}
\label{sec:discussion}
Our study shows that ML documentation artifacts such as ModelCards and DataSheets could contain substantial requirements-relevant information, meaning that they should be considered as key sources of requirements-relevant information as part of SE processes for ML systems. This validates their potential to serve as key boundary objects as part of communication and coordination in SE processes for ML systems~\cite{kasauli2020charting}. However, although abundant, we found that RE-relevant information is often redundant and inconsistently documented in this format. This inconsistency complicates reuse and traceability, which was also identified by Chang et al.'s~\cite{chang2022understanding} investigation of implementation challenges in ML documentation. \par

To understand the potential of current requirements representations in capturing this information we evaluated three RE templates and found each format had limitations: EARS lacks expressiveness for uncertainty, EARS and Rupp's template both struggle with documenting rationale and limitations; and Volere, though more expressive, uses requirements types too generic for ML-specific content. However, despite these limitations, these representations were reasonably successful in capturing RE-relevant information, particularly Volere, with it's capability to capture external factors. \par

Our main contribution is a systematic, bottom-up analysis of requirements information drawn from exemplary ML documentation. 
To our knowledge, this is the first study to examine ModelCards and DataSheets from an RE process perspective. 
Unlike prior work that proposes high-level frameworks~\cite{ahmad2023requirementsFramework,villamizar2024identifying} or targets specific requirement types~\cite{Bajraktari2024DocumentationON,habibullah2024framework,al2022resam}, our work quantified the potential presence of relevant information and evaluates how well established RE templates can capture real-world ML documentation content.
Our evaluation identifies key obstacles in bridging informal ML documentation artifacts and formal RE processes. 
Our results can help in supporting traceability: by establishing RE-guided processes for structuring component-level details from ML documentation artifacts, stakeholders can verify if a model or dataset meets broader system-level requirements, such as if a model's performance metrics align with system needs. \par
\textbf{Future Work}
\textit{For practitioners:} Our findings provide guidance for requirements engineers working with ML components. Our deductive criterion (Table~\ref{tab:REinfo}) can serve as a checklist to identify requirements relevant information in ML documentation for the RE process in ML systems development. Our evaluation also informs the selection of an appropriate RE language when writing requirements for ML components: EARS and Rupp's template are sufficient for capturing core functionality, while Volere is best suited for documenting context, limitations and dependencies. 
\textit{For researchers:} Volere's hierarchical structure suggests its potential as a template to mitigate the inconsistent levels of detail found in current ML documentation and reduce information redundancy. However, these established RE languages could also require adaptation for ML. Volere needs a more granular, ML-specific typology to distinguish concepts like data source constraints, evaluation metric requirements and data quality attributes (work on NFRs for ML~\cite{habibullah2024framework} might be a starting point for such additional concept distinctions), and EARS could be extended with flexible model keywords from Rupp's template (e.g., ``should'', ``may'') to better capture uncertainty. Our extraction method could be applied to other ML documentation types like Nutritional Labels~\cite{stoyanovich2019nutritional} and FactSheets~\cite{arnold2019factsheets} to generalize our findings. Our manually created dataset now provides a baseline to develop and evaluate tools that may automate requirement extraction from ML documentation using natural language processing or Large Language Models (LLMs).
\subsection{Threats to Validity}
\textit{Internal threats.} 
Although all data and analysis procedures are made available, some degree of subjectivity remains in interpreting whether documentation content qualifies as a requirement.
The use of templates demands a learning curve and understanding of the underlying semantics. 
While we applied a clear coding scheme and cross-checked 10\% of all decisions among authors, interpretive bias cannot be fully eliminated.\par
\noindent \textit{External threats.} 
Our document sample was selected using Google Scholar and sorted by default relevance, which may bias toward more cited or visible works. Our dataset could be seen as optimistic in terms of content and clarity due to extraction from highly-cited work, but our aim is only to evaluate the feasibility of requirements extraction from ML documentation, not to show that it will always work well in all cases.
Though we aimed for domain diversity in this dataset, some sectors may be underrepresented.
Our choice of requirements representation templates--EARS, Rupp's template, and Volere--was based on their prominence within the RE community~\cite{grosser2024benchmarking}. Other formats such as user stories or use cases could be evaluated in future similar studies. 

\section{Conclusion}
\label{sec:conclusion}
This study first showed that while ModelCards and DataSheets can be rich in requirements-relevant information, they are often descriptive, redundant, and inconsistently documented.
We showed further that three established RE process templates -- EARS, Rupp's template, and Volere -- are able to capture much of this information. 
However, constrained templates like EARS and Rupp fail to capture necessary ML-specific details like model limitations. 
Conversely, the Volere framework can capture this information but lacks domain-specific categories to classify it effectively. 
By identifying these gaps, our work lays a foundation for improving both RE and ML documentation processes. 

\begin{credits}
\subsubsection{\ackname} We received support from the Swedish Research Council (VR) iNFoRM Project and the Wallenberg AI, Autonomous Systems and Software Program (WASP).
\end{credits}

\bibliographystyle{splncs04}
\bibliography{sample-base}

\newpage
\bibliographystyleD{splncs04}
\sbox{\trashbox}{\parbox{\textwidth}{\bibliographyD{data}}}



\end{document}